# Electrochemical Degradation of Per- and Polyfluoroalkyl Substances (PFAS) using Low-cost Graphene Sponge Electrodes

Nick Duinslaeger[†,‡], Jelena Radjenovic [†,§,*]

[†]*Catalan Institute for Water Research (ICRA), Emili Grahit 101, 17003 Girona, Spain*

[‡]*University of Girona, Girona, Spain*

[§]*Catalan Institution for Research and Advanced Studies (ICREA), Passeig Lluís Companys 23, 08010 Barcelona, Spain*

*\* Corresponding author:*

*Jelena Radjenovic, Catalan Institute for Water Research (ICRA), Emili Grahit 101, 17003 Girona, Spain*

Phone: + 34 972 18 33 80; Fax: +34 972 18 32 48; E-mail: jradjenovic@icra.cat

# ABSTRACT

Boron-doped, graphene sponge anode was synthesized and applied for the electrochemical oxidation of C4-C8 per- and polyfluoroalkyl substances (PFASs). Removal efficiencies, obtained in low conductivity electrolyte (1 mS cm$^{-1}$) and one-pass flow-through mode, were in the range 16.7-67% at 230 A m$^{-2}$ of anodic current density, and with the energy consumption of 10.1 ± 0.7 kWh m$^{-3}$. Their removal was attributed to electrosorption (7.4-35%), and electrooxidation (9.3-32 %). Defluorination efficiencies of C4-C8 perfluoroalkyl sulfonates and acids were 8-24 % due to a fraction of PFAS being electrosorbed only at the anode surface. Yet, the recovery of fluoride was 74-87% relative to the electrooxidized fraction, suggesting that once the degradation of the PFAS is initiated, the C-F bond cleavage is very efficient. The nearly stoichiometric sulfate recoveries obtained for perfluoroalkyl sulfonates (91%-98%) relative to the electrooxidized fraction demonstrated an efficient cleavage of the sulfonate head-group. Adsorbable organic fluoride (AOF) analysis showed that the remaining partially defluorinated byproducts are electrosorbed at the graphene sponge anode during current application and are released into the solution after the current is switched off. This proof-of-concept study demonstrated that the developed graphene sponge anode is capable of C-F bond cleavage and defluorination of PFAS. Given that the graphene sponge anode is electrochemically inert towards chloride and does not form any chlorate and perchlorate even in brackish solutions, the developed material may unlock the electrochemical degradation of PFAS complex wastewaters and brines.

# INTRODUCTION

The global presence of per- and polyfluorinated substances (PFASs) in the environment is a topic of major concern due to their persistency, toxicity, and bio-accumulative potential (Niu et al., 2016; Ochoa-Herrera & Sierra-Alvarez, 2008; D. Zhang et al., 2016). The presence of multiple, highly stable carbon-fluorine bonds (bond dissociation energy up to 531.5 kJ mol$^{-1}$) (K. Zhang et al., 2013) and their amphiphilic nature make remediation of PFASs by means of oxidation or reduction extremely challenging. The addition of perflourooctanoic acid (PFOA), perfluorooctanesulfonate (PFOS) and their salts to the Annex B of the Stockholm Convention on Persistent Organic Pollutants (The Secretariat of the Stockholm Convention (SSC), 2019) has partially restricted the use of these chemicals since 2009. From 2019 onwards, PFOS was completely restricted under the EU POPs Regulation (EU Commission, 2019). Yet, both PFOS and PFOA are still frequently encountered in our water cycle (Podder et al., 2021).

Advanced oxidation processes based on ozone or hydroxyl radical ($^{\bullet}$OH) (e.g., UV/$H_2O_2$, Fenton-based processes) are ineffective in the defluorination of PFASs (Nzeribe et al., 2019). Non-destructive technologies like reverse osmosis (RO), ion exchange resins (IXR) and adsorption onto granular activated carbon (GAC) can remove efficiently long chain PFASs but result in a concentrated PFAS stream or residual that needs to be further treated. Electrochemical oxidation is a promising technology for the removal of PFAS from water due to its ability to achieve C-F bond breakage and complete defluorination (Radjenovic et al., 2020). To date, a range of anodes was shown to be capable of PFAS defluorination, such as mixed metal oxide (MMO) anodes based on $SnO_2$ and $PbO_2$ coatings (H Lin et al., 2012; Zhuo et al., 2011, 2020), boron-doped diamond (BDD) anode (Carter & Farrell, 2008; Liao & Farrell, 2009) and Magnéli phase titanium suboxide ceramic anode (Le et al., 2019; Lu Wang et al., 2020). However, all above-mentioned commercial anode materials produce free chlorine in the presence of chloride, which reacts rapidly with the organic matter to form persistent chlorinated organic by-products (Radjenovic & Sedlak, 2015). Furthermore, at high current densities typically required for PFASs defluorination, chloride is further oxidized to chlorate and perchlorate which pose a major challenge for the safe discharge of electrochemically treated water (Radjenovic et al., 2020).

In our recent studies (Baptista-Pires et al., 2021; Norra et al., 2021), we presented a new, graphene-based sponge electrode, which exhibited high electrocatalytic activity for electrochemical degradation of persistent organic and microbial contaminants, and remarkably low electrocatalytic activity for chloride oxidation; the current efficiency for chlorine production in the presence of 20 mM NaCl was only 0.04% at the 173 A m$^{-2}$ of anodic current density, and there was no chlorate and perchlorate formation (Baptista-Pires et al., 2021). The graphene-based sponges entailed other advantages such as their synthesis using a simple, low cost and easily scalable hydrothermal self-assembly method using mineral wool as supporting template, structural stability and flexibility, as well as the possibility to functionalize the graphene-based coating and tailor their electrocatalytic activity for the removal of persistent organic and microbial contaminants (e.g., iodinated contrast agents, triclosan, diclofenac, *Escherichia coli*) by a simple introduction of dopants (e.g., atomic dopants, two-dimensional materials, MXenes) (Baptista-Pires et al., 2021; Norra et al., 2021). Moreover, the material demonstrated exceptional stability to anodic polarization, likely due to the strong interfacial adhesion and covalent bonding between the graphene nanosheets and $SiO_2$ (Hintze et al., 2016), a major component of the mineral wool supporting template.

In this study, we report on the electrochemical oxidation of six model PFASs, i.e., perfluorooctanesulfonic acid (PFOS), perfluorooctanoic acid (PFOA), perfluorohexane sulfonate (PFHxS), perfluorohexanoic acid (PFHxA), perfluorobutanesulfonic acid (PFBS) and perfluorobutanoic acid (PFBA) using graphene-based sponge anode. Electrochemical system equipped with boron-doped reduced graphene oxide (BRGO)-coated sponge anode and stainless-steel sponge cathode was operated in flow-through, one-pass mode. B-doping was previously observed to increase the electrocatalytic activity and yield significantly lower resistance compared to the undoped graphene sponge electrode (Baptista-Pires et al., 2021; Wu et al., 2016). We evaluated the impact of the applied anodic current and the electrolyte flow rate on the electrochemical removal of PFASs. All experiments were conducted using a low-conductivity supporting electrolyte (1 mS cm$^{-1}$) to evaluate the system performance under the ohmic drop of real contaminated water. The amount of the released fluoride and, in the case of perfluorosulfonic acids (PFSAs), sulfate, was measured to determine the % of electrochemical degradation of PFAS. The obtained results indicate that the developed graphene sponge anode can electrochemically degrade C4-C8 PFAS and, given its previously demonstrated low electrocatalytic activity for chloride oxidation, this can be achieved without forming toxic and persistent organic and inorganic chlorinated by-products.

**MATERIALS AND METHODS**

**Materials**

All chemicals used in the experiments were reagent grade or higher and used as received. Graphene oxide (GO) aqueous dispersion (4 g L$^{-1}$) was provided from Graphenea S.L. PFOS (CAS 1763-23-1), PFOA (CAS 335-67-1), PFHxS (CAS 355-46-4), PFHxA (CAS 307-24-4), PFBS (CAS 375-73-5), PFBA (CAS 375-22-4) were purchased from Sigma-Aldrich Chemical. Urea, boric acid and phosphate salts were purchased from Sigma Aldrich. Mineral wool was procured from Diaterm. All solutions were prepared in Milli-

Q water with a conductance of 18.2 MΩ cm at 25 ± 1 °C. High-purity methanol used for mass spectrometry analysis was purchased from Fisher Scientific.

**Graphene Sponge Synthesis**

Graphene-based sponges were synthesized according to the previously reported procedure (Norra et al., 2021). In brief, to synthesize a boron-doped reduced graphene oxide anode (BRGO), 6 g of boric acid was dissolved in 2 g L$^{-1}$ GO. The template for the graphene sponge synthesis was mineral wool, which was submerged in the GO/boric acid solution and successively squeezed to ensure it was completely soaked. The soaked sponge was subjected to hydrothermal synthesis for 12 h at 180 °C. The obtained B-doped graphene-based sponge was then soaked in MilliQ water to remove the unbonded graphene and the remaining impurities. As a final step, the sponges were cut to the right size before being employed as anode for the electrooxidation of PFASs. Besides the high chemical, thermal and structural stability of the mineral wool template, graphene sponge electrodes produced using this method present high electrochemical stability during both anodic and cathodic polarization, due to the formation of C-Si and C-O covalent bonds between the silicate present in the mineral wool and graphene, and thus strong interfacial adhesion between the RGO nanosheets and SiO$_2$ (Hintze et al., 2016; Ramezanzadeh et al., 2016; Shemella & Nayak, 2009). The characterization of the graphene sponges used in this study was reported in our recent work (Baptista-Pires et al., 2021; Norra et al., 2021), and is summarized in **Text S1**.

**Electrochemical Oxidations Experiments**

Electrochemical oxidation experiments were performed in a cylindrical flow-through reactor (diameter 5 cm, projected surface area 17.34 cm$^2$). The reactor was operated in one-pass, continuous mode at flow rates of 2.5, 5, and 10 mL min$^{-1}$ using a digital gear pump (Cole-Palmer), resulting in hydraulic residence times (HRTs) of 7, 3.5 and 1.7 min, and effluent fluxes of 86.5, 173 and 346 L m$^{-2}$ h$^{-1}$ (LMH), respectively. The graphene sponge anode of approximate thickness 0.5 cm was fed with current using a three dimensional (3D) stainless steel current collector (Baptista-Pires et al., 2021), whereas the stainless steel sponge cathode (Spontex) was fed with the current directly. Physico-chemical properties of the target PFASs, including the molecular polarizabilities of the PFAS anions, as predicted using the density functional theory (DFT) modelling (Radjenovic et al., 2020), are summarized in **Table S1**. At the start of each experiment, initial open circuit (OC$_0$) runs were conducted to evaluate the possible losses of the target PFAS through adsorption onto the graphene sponge electrodes. At the end of each chronopotentiometric experiments, the effluent was sampled after the current was switched off (i.e., OC$_{final}$) to evaluate the overall loss of the target PFASs due to electrosorption only (i.e., without subsequent degradation). Before starting each experiment, the reactor was flushed with the supporting electrolyte solution (10 mM phosphate buffer). To determine the impact of the anodic current on the PFASs removal, experiments were conducted at varying flowrates in the chronopotentiometric mode at 29, 58, and 115 A m$^{-2}$ of applied anodic current density (calculated using the projected anode surface area) using a BioLogic multi-channel potentiostat/galvanostat VMP-300 and a leak-free Ag/AgCl reference electrode (Harvard Apparatus). Electrochemical impedance spectroscopy (EIS) was used to determine the ohmic internal resistance and ohmic drop at each applied current, whereas the EIS experimental data was fitted using the BioLogic EC-lab software. All electrode potentials were expressed versus SHE (/SHE, V). Samples were taken after 10, 20 and 30 bed volumes at each current. The six model PFASs, i.e., PFOS, PFOA, PFHxS, PFHxA, PFBS and PFBA, were added to 10 mM phosphate buffer (Na$_2$HPO$_4$/NaH$_2$PO$_4$, pH 7.2, 1 mS cm$^{-1}$) at 0.2 μM initial concentration. The pH remained constant at pH 7-7.2 in all experiments. Samples were collected in polypropylene vials to avoid trace contamination with PFASs. In like manner, use of glassware was avoided, where possible, to prevent losses due to adsorption of PFASs on glass surfaces.

**Determination of defluorination and desulfonation of PFAS**

To allow the detection of fluoride released by the C-F bond cleavage, experiments were performed at 230 A m$^{-2}$ of applied anodic current density and with each target PFAS separately at 2 µM initial concentration., during 30 bed volumes. To determine the portion of the adsorbed, electrosorbed, electrochemically degraded and volatilized PFAS, the experiments were conducted as follows.

First, removal due to volatilization was determined for each PFAS by using a closed effluent reservoir container connected to the alkaline trap with 100 mM NaOH solution, as illustrated in **Figure S1**. The alkaline solution was subjected to target PFASs analysis, as well as adsorbable organic fluorine (AOF) analysis -representing fluorinated organics, using a previously published method based on combustion ion chromatography (CIC) (Wagner et al., 2013). For AOF analysis, the alkaline solution was acidified with nitric acid to improve the adsorption of the organically bound fluor to the activated carbon columns.

Next, effluent samples were collected as cumulative fractions in the experimental runs consisting of OC$_0$ - 230 A m$^{-2}$ - OC$_{final}$ sequences, to measure the entire amount of the target PFASs remaining in the effluent, as well as AOF representing target PFASs and any partially fluorinated byproducts present in the effluent. To determine the amount of the fluoride released from the PFASs, three different approaches were employed: *i)* analysis of the fluoride present in the reactor effluent using ion chromatography (IC), *ii)* washing of the employed BRGO sponge with 100 mM NaOH to release any electrosorbed fluoride via ion exchange, and analysis of the washing solution using a fluoride probe

(Orion Star, Thermo Scientific) *iii)* CIC analysis of the BRGO anode, as described further in the text. In addition, to detect fluoride released from PFASs and bound in the graphene coating via X-ray photoelectron spectroscopy (XPS) analysis, long-term electrochemical degradation experiment (i.e., 800 bed volumes) was conducted at a high initial PFASs concentration (2 µM) and at 230 A m$^{-2}$, to increase the presence of bound F in the BRGO anode. Prior to the analysis of fluoride bound in the coating, the employed BRGO anode was thoroughly washed with MilliQ water and a 100 mM NaOH aqueous solution to remove any adsorbed fluorinated organics. To verify further that there are no adsorbed fluorinated organics interfering in the CIC analysis of fluoride, additional experiments were performed with PFOA and PFOS, and the employed graphene sponge anodes were washed with 100 mM ammonium acetate in methanol, described by (Xiao et al., 2020) as the optimum solution to extract PFAS from carbonaceous materials. The washing solution was then analyzed for the presence of the target PFASs as well as for AOF.

The CIC analysis of the fluoride incorporated in the graphene sponge anode was performed by combusting the used graphene sponge and measuring the amount of the released fluoride. For this purpose, freshly synthesized BRGO sponge anode was employed for electrochemical degradation of each PFAS separately. The used BRGO anode was analyzed using CIC in a similar manner to the analysis of loaded activated carbon cartridges employed in the AOF analysis, i.e., by combusting the sponge at 1,100ºC, trapping the HF gas released from the sponge in the absorbent solution, and quantifying it as fluoride using ion chromatography (IC). This analysis was performed after each experiment with single PFASs and using a newly synthesized graphene sponge for each run. As previously mentioned, to ensure that the measured fluoride represents fluoride incorporated into the graphene coating and not fluoride originating from the adsorbed PFASs or their byproducts, the sponges were thoroughly washed with MilliQ water and 100 mM NaOH solution. Also, previous study revealed the potential presence of fluorine in inorganic thermal insulation materials such as mineral wool template employed in the graphene sponge synthesis (Liu et al., 2018). Thus, CIC method was employed to determine the fluoride in the mass of mineral wool template employed in the graphene-based sponge synthesis. The amount of the fluoride incorporated into the BRGO anode due to PFAS degradation was obtained by subtracting the fluoride released from the mineral wool from the amount detected after combusting the graphene sponge used for electrochemical degradation of a specific PFAS.

To determine the release of sulfate from the PFSAs, the reactor was operated with the polarity switch conducted after the experimental run with the target PFSAs. The BRGO anode was polarized to -1 V/SHE, and the retained sulfate was recovered as a concentrated aqueous solution. All experiments were conducted at least in duplicate, and the results are expressed as mean with their standard deviations (SDs).

**Analytical Methods**

Target PFASs were analyzed using a 5500 QTRAP hybrid triple quadrupole-linear ion trap mass spectrometer (QqLIT-MS) with a turbo Ion Spray source (Applied Biosystems, Foster City, CA, USA), coupled to a Waters Acquity Ultra-Performance$^{TM}$ liquid chromatograph (UPLC, Milford, MA, USA). The analytical method for PFASs analysis was adapted from (Llorca et al., 2012) and is described in detail in **Text S3**. The concentration of sulfate and fluoride was determined using high-pressure IC (HPIC, Dionex ICS-5000 system). Fluoride in highly alkaline solutions (i.e., 100 mM NaOH) was analyzed using a selective fluoride probe (Orion Star, Thermo Scientific). Determination of AOF in the aqueous samples was performed with an AQF-2100H Automatic Quick Furnace for CIC (Mitsubishi Chemical Analytech), coupled to IC (ICS-

2100, Dionex). The analytical method for AOF analysis was adapted from (Wagner et al., 2013) and is described in the **Text S4**. XPS measurements were done with a Phoibos 150 analyzer (SPECS GmbH, Berlin, Germany) in ultra-high vacuum conditions (base pressure 1-10 mbar) with a monochromatic aluminium Kalpha x-ray source (1486.74 eV). The energy resolution as measured by the FWHM of the Ag 3d5/2 peak for a sputtered silver foil was 0.58 eV.

## RESULTS AND DISCUSSION

**Impact of the anodic potential and flow rate on PFASs removal**

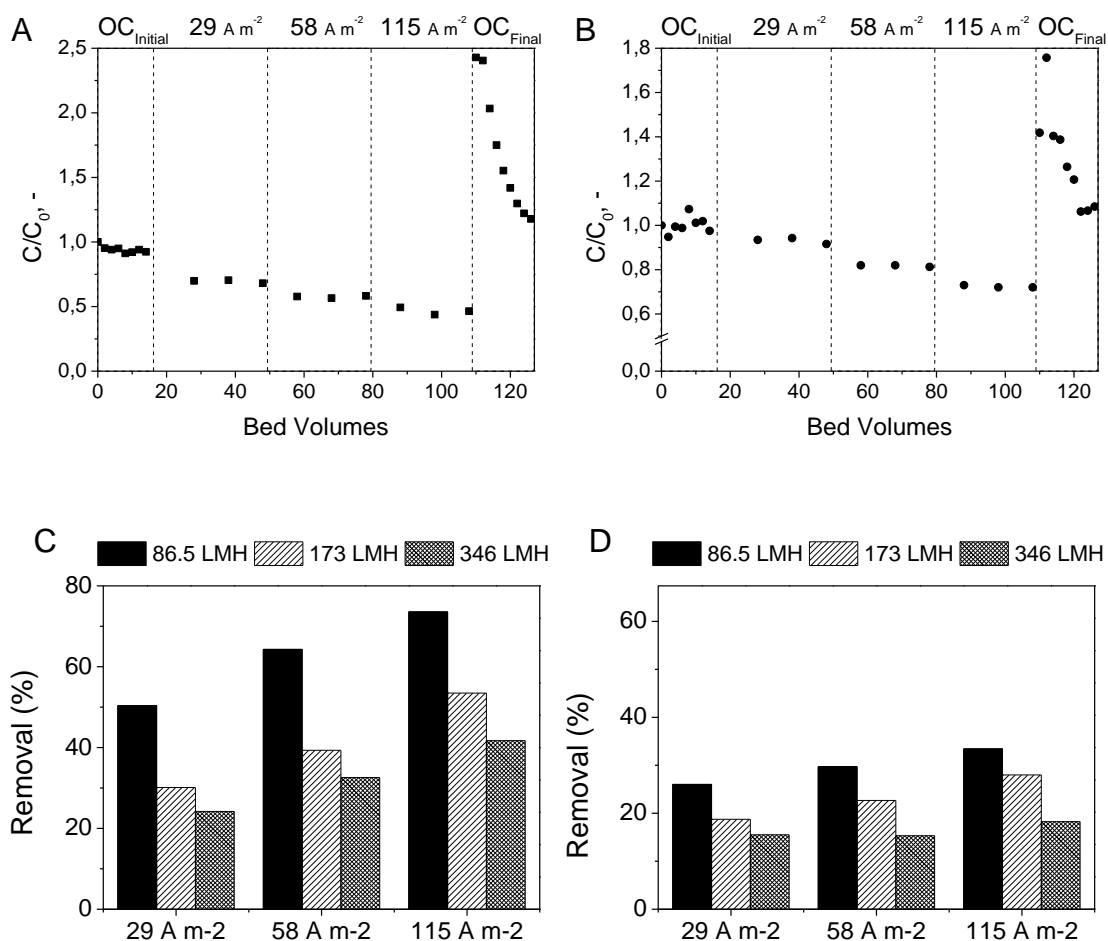

**Figure 1**. Measured concentrations of **A)** PFOS and **B)** PFOA normalized to the initial value (C/C$_0$, C$_0$=0.2 µM) at varying anodic currents of 29, 58 and 115 A m$^{-2}$ and using a flow rate of 5 mL min$^{-1}$ (i.e., 173 LMH), and overall removal of **C)** PFOS and **D)** PFOA at varying anodic currents (29, 58 and 115 A m$^{-2}$) and flowrates (2.5, 5 and 10 mL min$^{-1}$, i.e., 86.5, 173 and 346 LMH).

**Figure 1** illustrates the observed removals of long-chain PFOS and PFOA at varying current densities and flowrates, whereas the results of all six PFASs are represented in **Figure S2**. None of the target PFASs was adsorbed onto the graphene-based sponge anode (or other reactor components), and the effluent concentrations were equal to their influent concentrations in the $OC_0$ run. Graphene-based sponge is a highly hydrophobic material with the contact angle of 139.67°±4.50° determined for the BRGO (Baptista-Pires et al., 2021). Yet, the hydrophobic fluorinated tail of PFASs did not interact with the unpolarized graphene sponge. This may be a consequence of the negative surface charge of the BRGO sponge at pH 7, as the determined zeta potential at this pH value is –36.2 mV (Norra et al., 2021). Thus, electrostatic repulsion with the PFASs anions (pKa= -3.94 – 2.9, **Table S1**) plays a dominant role.

In the chronopotentiometric experiments conducted at 29, 58, and 115 A m$^{-2}$ of the applied anodic current density, the ohmic drop corrected anode potentials were 3.4, 4.7 and 6.1 V/SHE, respectively. The obtained Nyquist plot and ohmic-drop correction of the anode potentials is explained in **Text S2**. Application of 29 A m$^{-2}$ resulted in 30.5±1.2% (PFOS), 6.9±1.4% (PFOA), 16±5.1% (PFHxS), 9.4±1.7% (PFHxA), 12.2±1.0% (PFBS) and 10.5±0.3% (PFBA) removal of the target PFASs (**Figure S2**). Further increase in the current density to 58 and 115 A m$^{-2}$ led to moderate increases in the removal efficiencies, with 53.5±2.8% (PFOS), 27.7±0.6% (PFOA), 35.3±1.8% (PFHxS), 19.8±1% (PFHxA), 20.1±1.6% (PFBS) and 16.5±1.6% (PFBA) removal at 115 A m$^{-2}$. Given that higher currents also enhanced water electrolysis, increase in the PFAS electrosorption and electrochemical degradation at higher currents was limited by the more pronounced evolution of oxygen bubbles from the BRGO anode. In all experiments, the flow-through reactor was in steady state after the first 10 bed volumes (**Figure S2**). The removal efficiencies for both PFSAs and perfluorocarboxylic acids (PFCAs) decreased with the shortening of the chain length in the order PFOS>PFHxS>PFBS. The chain length-dependent reactivity of PFASs in electrochemical oxidation was previously observed for other anode materials such as BDD (Trautmann et al., 2015), Ce-doped $PbO_2$ (Niu et al., 2012) and $Ti_4O_7$ (Y. Wang et al., 2019). Also, at all applied current densities, the PFSAs demonstrated removal efficiencies higher than those of the PFCAs for the same perfluoroalkyl chain lengths. For example, at 29, 58 and 115 A m$^{-2}$ and 173 LMH of applied effluent flux, the removal efficiencies were 30.6±1.2, 39.3±2.3% and 53.5±2.8% for PFOS, and 6.9±1.4, 18.3±0.4, 27.7±0.6 % for PFOA, respectively (**Figure 1**). Although previous studies report overall higher electrochemical removal of PFCAs compared with PFSAs (Le et al., 2019; Schaefer et al., 2017; Zhuo et al., 2012), some authors have reported the opposite trend (Schaefer et al., 2015; Y. Wang et al., 2019). This may be a consequence of lower molecular polarizabilities ($α_p$) of the PFSA anions compared with the PFCA anions (Radjenovic et al., 2020). Molecular polarizability predicted using the density functional theory (DFT) modelling describes the ability of a compound to form a dipole in response to an external electric field, and was determined to be the only molecular descriptor of PFASs that correlated with their experimental rates of electron transfer reported in literature (Radjenovic et al., 2020). For PFSAs, molecular polarizabilities decrease in the order PFOS (170 C m$^2$ V$^{-1}$)>PFHxS (138 C m$^2$ V$^{-1}$)>PFBS (107 C m$^2$ V$^{-1}$) (**Table S1**). In the case of PFCAs, lower $α_p$ values are predicted, i.e., PFOA (140 C m$^2$ V$^{-1}$)>PFHxA (109 C m$^2$ V$^{-1}$)>PFBA (77.5 C m$^2$ V$^{-1}$) (**Table S1**), which explains their overall lower removal observed in our experiments. In the OC$_{final}$, the effluent concentrations for all target PFASs rose above their initial concentrations (**Figure 1**, **Figure S2**), indicating that a portion of the removed PFASs was merely electrosorbed onto the graphene sponge anode and released when the application of current was stopped. This was further investigated in the experiments conducted with single PFAS solutions, as explained further in the text.

Applied flow rate impacted the observed removal of PFASs, with worsened performance at higher applied effluent fluxes of 346 LMH (**Figure S3**). Lowering of the effluent flux from 173 to 86.5 LMH significantly improved the removal efficiencies of PFSAs to 73.5±1.9% (PFOS), 49.1±1.6% (PFHxS), and 39.9±1.5% (PFBS) at 115 A m$^{-2}$. In the case of PFCAs, the increase in removal % was somewhat less pronounced, with 33.4±1.6% (PFOA), 25.7±2.7% (PFHxA), 30.6±0.6% (PFBA) removal at 86.5 LMH, in line with their overall lower molecular polarizabilities. Thus, electrochemical removal of PFASs at the graphene-based sponge anode was limited by their mass transfer and diffusion towards the electrode surface. The average volumetric pore size of the graphene sponge anode is 10-50 μm (Baptista-Pires et al., 2021), thus the convection-enhanced mass transfer of trace pollutants cannot be achieved to the same degree as in the case of carbon nanotube (CNT)-based electrodes and porous $Ti_4O_7$ electrodes with very small pore diameters (Trellu et al., 2018). For treating larger volumes of water without compromising the electrochemical removal of PFASs, several modules equipped with the graphene-based sponge anodes and larger anode sizes (e.g., 10x10 cm) should be employed. The estimated price of the graphene-based sponge is only €23-46 per m$^2$ (Baptista-Pires et al., 2021; Norra et al., 2021), extremely competitive compared with the state-of-the-art BDD anode (~€4,200-6,000 per m$^2$) and dimensionally stable anodes (DSAs) (€3,000 per m$^2$) (Wenderich et al., 2021), and is thus not considered a limitation for the construction of an electrochemical unit with large specific anode surface area. Nevertheless, energy consumption for the treatment of real contaminated water should be carefully evaluated and applied current densities and other operating parameters (e.g., flowrate) need to be optimized to achieve complete PFASs removal and defluorination in the presence of organic matrix and inorganics (e.g., chloride, carbonate).

**Electrosorption and electrochemical degradation of the PFASs**

To determine the portions of volatilized, electrosorbed, and degraded PFASs, as well as the efficiencies of the C-F bond cleavage and desulfonation, the experiments were conducted using single solutions of PFASs at 2 µM initial concentration and anodic current density of 230 A m$^{-2}$, resulting in the ohmic-drop corrected anode potential of 7.6 V/SHE. The alkaline trap employed in the preliminary experiments to evaluate the % of PFAS volatilization did not contain any of the target PFASs, and the AOF values were below the detection limit (i.e., 0.5 µg L$^{-1}$), implying that there was no volatilization of the target PFASs or their byproducts from the reactor. The amount of the electrosorbed PFASs was calculated based on their concentrations measured in the cumulative fractions sampled in the OC$_{final}$ (**Text S4**). The % of the electrooxidation of the target PFAS was calculated by subtracting the amount of the PFAS recovered during the OC$_{final}$ (i.e., corresponding only to the electrosorbed PFAS) from the total amount removed during current application, and dividing it with the latter (**Text S5**). The total removal efficiencies presented as removal % due to electrosorption, and removal % achieved by electrooxidation are presented in **Figure 2A**. As expected, the longer chain PFASs demonstrate the highest electrooxidation efficiency of 32±4.3% and 31.5±9.9% for PFOS and PFOA, respectively. Electrosorbed (but not degraded) fraction of PFOS was significantly higher (35±8.3% removal due to electrosorption) compared with the PFOA (12.1±6.5% removal), as expected based on their predicted molecular polarizabilities (**Table S1**). The percentages of electrosorption as well as electrooxidation further decrease progressively in the order C8>C6>C4 for both PFSAs and PFCAs, with the overall lower removal % observed in the latter case (**Figure 2A**).

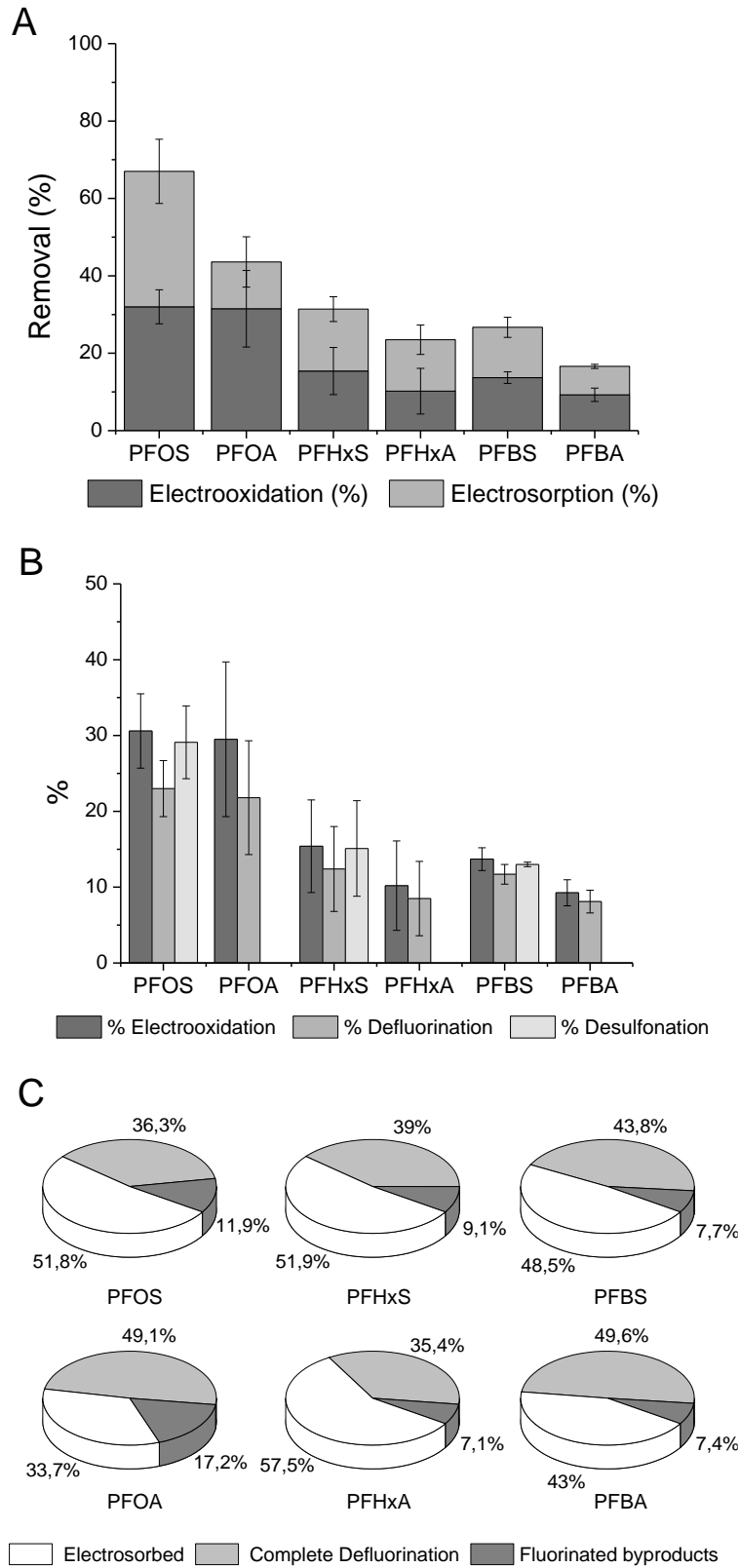

**Figure 2**. **A)** Removal efficiency of the target PFASs, represented as a sum of removal due to electrooxidation and electrosorption, **B)** electrooxidation, defluorination and desulfonation (where applicable) efficiency of the target PFASs, and **C)** distribution of different removal mechanisms in the overall removal of each PFAS. The experiments were conducted at 230 A m$^{-2}$ and 2 μM initial concentration of each compound.

**Table 1.** Removal efficiencies of PFASs (230 A m$^{-2}$, C$_0$=2 µM), with calculated values for electrosorption, volatilization, electrooxidation, defluorination and desulfonation, as well as the relative % of the measured AOF.

| Compound | Removal (%) | Electrosorption (%) | Volatilization (%) | Electrooxidation (%) | Defluorination (%) | Desulfonation (%) | Partially defluorinated intermediates (%) |
|---|---|---|---|---|---|---|---|
| PFOS | 67 ± 4.1 | 35.0 ± 8.3 | <LOQ* | 32 ± 4.4 | 24 ± 3.3 | 29.1 ± 4 | 8 ± 1.1 |
| PFOA | 43.6 ± 12.1 | 12 ± 6.5 | <LOQ* | 31.5 ± 9.9 | 21.8 ± 7.5 | NA | 9.7 ± 3 |
| PFHxS | 31.3 ± 2.9 | 16.0 ± 3.2 | <LOQ* | 15.4 ± 6.1 | 12.4 ± 4.9 | 15.1 ± 6 | 3 ± 1.1 |
| PFHxA | 23.5 ± 2.1 | 13.3 ± 3.8 | <LOQ* | 10.2 ± 5.9 | 8.5 ± 4.9 | NA | 1.7 ± 0.9 |
| PFBS | 26.7 ± 1.2 | 13.0 ± 2.6 | <LOQ* | 13.7 ± 1.5 | 11.7 ± 1.3 | 13 ± 1.4 | 2 ± 0.2 |
| PFBA | 16.6 ± 1.2 | 7.4 ± 0.5 | <LOQ* | 9.3 ± 1.7 | 8.1 ± 1.5 | NA | 1.2 ± 0.2 |

For the shorter-chain PFASs, removal due to electrooxidation was 15.3±6.1% (PFHxS), 10.2±5.9% (PFHxA), 13.7±1.5% (PFBS), and 9.3±1.7% (PFBA). Removal due to electrosorption only was calculated to be 35±8.3% (PFOS), 12.6±6.5% (PFOA), 16±3.2% (PFHxS), 13.3±3.8% (PFHxA), 13±2.6% (PFBS) and 7.4±0.5% (PFBA). The observed electrosorption of PFASs without their further electrooxidation is likely a consequence of the low supporting electrolyte conductivity that limits the electrocatalytic activity of the electrode (Radjenovic et al., 2020).

Analysis of sulfate in the reactor effluent did not show any release of sulfate during the application of current. However, sulfate was detected in the reactor effluent after switching off the current, although in the amounts lower than expected based on the obtained electrooxidation % (**Figure 2B**). To evaluate the desulfonation % of the target PFSAs, separate experiments were conducted with the BRGO anode employed for PFSAs degradation polarized after each experiment to -1 V/SHE to desorb the retained $SO_4^{2-}$. The % of desulfonation was calculated based on the total sulfate measured, and the amount of the theoretically expected sulfate was calculated from the electrooxidation removal % (**Text S6**). The recoveries of $SO_4^{2-}$ obtained for PFOS, PFHxS and PFBS were 91%, 98% and 95%, respectively. Thus, nearly stoichiometric sulfate recovery demonstrated that electrooxidized PFASs underwent cleavage of the sulfonate head-group. The desulfonation was 29.1±4.8% (PFOS), 15.1±6.3% (PFHxS) and 13.0±0.3% (PFBS) (**Figure 2B, Table 1**), calculated considering the initial concentration of the target PFASs (**Text S6**). The incomplete mass balance for sulfate in the case of PFOS could be attributed to recombination of the cleaved $SO_4^{2-}$ with $C_nF_{2n-1}^{\bullet}$ radicals (Yang et al., 2013).

The initial release of sulfate was previously reported for electrooxidation of PFSAs, and is achieved in the first electron transfer step to the anode (eq. 1) followed by a Kolbe desulfonation to produce $C_nF_{2n-1}^{\bullet}$ (eq. 2) (Carter & Farrell, 2008; Niu et al., 2013).:

$$C_nF_{2n+1}SO_3^- \rightarrow C_nF_{2n+1}SO_3\cdot + e^- \qquad \text{(eq. 1)}$$

$$C_nF_{2n+1}SO_3\cdot + H_2O \rightarrow C_nF_{2n+1}\cdot + SO_4^{2-} + 2H^+ \qquad \text{(eq. 2)}$$

Analogous electron transfer reaction occurs for PFCAs (eq. 3) followed by a Kolbe decarboxylation (eq. 4):

$$C_nF_{2n+1}COO^- \rightarrow C_nF_{2n+1}COO\cdot + e^- \qquad \text{(eq. 3)}$$

$$C_nF_{2n+1}COO\cdot \rightarrow C_nF_{2n+1}\cdot + CO_2 \qquad \text{(eq. 4)}$$

According to the DFT simulations, the formed $C_nF_{2n-1}^\bullet$ radical reacts further with $OH^\bullet$, $O_2$, and $H_2O$ and undergoes $CF_2$ unzipping reaction mechanism to form HF, $CO_2$, and defluorinated byproducts (eq. 5-7), with the initial direct electron transfer being the rate-limiting step (Niu et al., 2013). This mechanism was supported by the DFT simulations (Carter & Farrell, 2008; Liao & Farrell, 2009) and experimental evidence that shows limited reactivity of PFASs to hydroxyl radical ($^\bullet OH$) attack (Trojanowicz et al., 2018).

$$C_nF_{2n+1}\cdot + HO\cdot \rightarrow C_nF_{2n+1}OH \qquad \text{(eq. 5)}$$

$$C_nF_{2n+1}OH \rightarrow C_nF_{2n-1}OF + HF \qquad \text{(eq. 6)}$$

$$C_nF_{2n-1}OF + H_2O \rightarrow C_{n-1}F_{2(n-1)+1}COO^- + HF + H^+ \qquad \text{(eq. 7)}$$

Fluoride released from the C-F bond cleavage could not be detected in the effluent samples during current application, in the $OC_{final}$, nor with the polarity reversal employed to recover the released sulfate. Previous studies (Guan et al., 2007; Hori et al., 2006; Hui Lin et al., 2013) indicated that fluorine or fluorine-containing species can remain present at the anode surface. An attempt to recover fluoride from the BRGO anode employed for the PFAS degradation by washing of the sponge with 100 mM NaOH solution did not result in the fluoride recovery via ion exchange. Furthermore, there was no AOF of PFASs measured in the washing solution, confirming that the sponge did not adsorb the target PFASs or any byproducts of their electrooxidation, in accordance with the data obtained in the $OC_0$ runs. Given that the HF released during PFAS defluorination is a highly reactive species, it was likely that fluoride was incorporated into the graphene coating of the BRGO anode, as shown in Figure S6. For example, HF is used for the synthesis of fluorinated graphene for optoelectronics and photonics applications and results in the formation of covalent C-F bonds with graphene (Ho et al., 2014; Liping Wang et al., 2014; Withers et al., 2010). Also, recent study (Y. Wang et al., 2022) reported the formation of a Ti-F bond in $Ti_4O_7$ anode after the electrochemical oxidation of a PFOS solution. XPS analysis of the BRGO anode used in a long-term experiment (800 bed volumes) at a high initial PFASs concentration (2 µM) did not allow the detection of fluoride, likely due to the very low relative % of the incorporated F. Nevertheless, fluoride incorporated into the BRGO anode could be detected using a CIC method. Preceding the CIC analysis, the BRGO anodes were washed with either 100 mM NaOH aqueous solution or 100 mM ammonium acetate in methanol to eliminate any fluorinated organics. The impact of both washing solutions (100 mM NaOH aqueous solution and 100 mM ammonium acetate in methanol) on the $F^-$ recovery was evaluated for the two PFAS compounds most likely to adsorb, PFOS and PFOA (Gagliano et al., 2020). Similar results in terms of $F^-$ recovery were obtained for both washing solutions (**Figure S5**), which is in line with the absence of PFASs adsorption in the $OC_0$ at the start of the experiments suggesting that adsorption is probably not the major removal mechanism. The rest of the discussed results were obtained by washing with a 100 mM NaOH aqueous solution.

The BRGO anodes employed for the degradation of single PFAS solutions were subjected to CIC analyses, and the $F^-$ concentrations obtained in the combustion of each sponge were used to calculate the fluoride recoveries for each PFASs (**Text S7**). The % of fluoride recovered from the amount of the electrooxidized PFASs were 75.3% (PFOS), 74.1% (PFOA), 81.1% (PFHxS), 83.3% (PFHxA), 85% (PFBS) and 87.1% (PFBA). Relative to the overall removal of PFASs, the overall defluorination ratios ranged from 24.0±3.3% (PFOS), 21.8±7.5% (PFOA), 14.5±4.9% (PFHxS), 8.5±4.9% (PFHxA), 11.7±1.3% (PFBS) to 8.1±1.5% (PFBA). More efficient overall C-F bond breakage for longer-chain PFOS and PFOA is in accordance with the previous studies (Schaefer et al., 2015; Y. Wang et al., 2019). However, this is likely another consequence of their more pronounced interaction with the graphene-based sponge anode surface compared with the shorter-chain PFASs, as the amount of the recovered fluoride relative to the amount of the electrooxidized compound did not exhibit the same decreasing tendency with the chain shortening, and was in the range of 74-87%, as explained above. The recovery of the bound fluoride from the BRGO sponge evidences the capability of the BRGO anode to cleave the C-F bond in both long-chain (C8) as well as short-chain (C4) PFASs. Although the observed defluorination % were low due to a significant portion of the target PFASs only undergoing electrosorption (i.e., without further degradation), it is important to note that) these results were obtained in a low conductivity solution (1 mS cm$^{-1}$). Furthermore, graphene sponge anodes may enable efficient C-F bond cleavage without compromising the quality of the treated water, as there is no formation of chlorate and perchlorate even in the presence of high chloride concentrations, and the current efficiency for chlorine formation was only 0.04% in the presence of 20 mM NaCl (Baptista-Pires et al., 2021; Norra et al., 2021). This makes it a unique anode material. Nevertheless, the impact of Cl$^-$ and other inorganic ions (e.g., sulfate, carbonate) on the electrochemical defluorination of PFASs needs to be further investigated. Thus, graphene-based sponge anodes may be well suited for the treatment of highly saline, more conductive streams rich in PFASs (e.g., reverse osmosis (RO) concentrate, landfill leachate), and avoid the increase in the effluent toxicity typically observed in the electrooxidation of highly saline streams (Radjenovic et al., 2011).

Fluorine modifies the electronic properties of graphene oxide by introducing scattering centres, widening the band gaps and reducing the charge in the conducting $\pi$ orbitals (Feng et al., 2016; Meduri et al., 2013; Robinson et al., 2010). Fully fluorinated graphene oxide exhibits a reduction of electric conductivity. Although the incorporation of fluoride into the BRGO anode did not change its electrochemical activity towards PFASs removal during the previously described long-term experiment (i.e., 800 bed volumes) at high PFASs concentration (mixed solution of six target PFASs at $C_0=2$ μM), further work is needed to evaluate long-term performance of the system when treating concentrated PFASs solutions. The bound F was present in traces in the graphene coating as it was not detectable in the XPS analyses and could only be determined by the complete combustion of the sponge and CIC analyses. If the graphene sponge anode performance becomes limited by the fluoride incorporation from PFASs degradation, the electrode can be easily replaced with a new graphene sponge considering its extremely low cost (€23-46 per $m^2$ of the projected surface area) (Baptista-Pires et al., 2021; Norra et al., 2021).

Given that not all fluoride could be recovered relative to the amount of the electrooxidized PFAS (i.e., 74-87% defluorination relative to the electrooxidized compound, **Figure 2C**), their electrochemical defluorination was incomplete. To estimate the amount of the formed partially defluorinated byproducts, AOF of the cumulative effluent samples was measured during the chronopotentiometric run at 230 A $m^{-2}$ and in $OC_{final}$, and the total amounts determined are presented in **Figure 3**. The measured AOF values are expressed in quantity of $F^-$ and presented versus the amount of the AOF expected based on the measured concentrations of each PFASs. During the application of current, the measured effluent AOF is comparable to the AOF expected from the remaining amount of the parent compound in the case of all six PFASs, indicating that the measured organically bound fluorine can be completely assigned to the parent compound. However, significant differences between the measured and expected AOF can be observed after the current is switched off, in the $OC_{final}$. For PFOS, the cumulative organic fluorine in the $OC_{final}$ expected from the PFOS measured in the effluent was 173.5 µg $F^-$, while the AOF analysis detected a total of 324.5 µg $F^-$. This higher measured AOF than expected AOF indicates the presence of an organic source of fluorine different from the parent compound. Similar to the desorption of the target PFASs from the BRGO anode in the $OC_{final}$ (**Figure 1**), any partially defluorinated byproducts underwent the same process, being strongly electrosorbed during the application of current, and released after the current was switched off. Thus, their accumulation at the anode during current application and simultaneous release in $OC_{final}$ leads to AOF values higher than expected based on the effluent concentrations of the parent compound. It should be noted here that the AOF is a semi-quantitative, bulk measurement because it is not possible to determine the exact extraction recoveries of the unknown fluorinated byproducts formed in electrooxidation of PFASs. Based on the defluorination % obtained relative to their overall % of electrochemical degradation, the AOF likely corresponds to short-chain, partially defluorinated byproducts (Niu et al., 2012). Moreover, it is likely that the measured AOF in the $OC_{final}$ overestimates the amount of the present short-chain, partially fluorinated degradation products considering that the recoveries of the AOF analyses of the target PFASs increase with fluorinated chain shortening, in the order PFOS<PFHxS<PFBS, and PFOA<PFHxA<PFBA (**Table S2**). The detection of the organically bound fluorine different from the parent PFASs after switching off the current further demonstrates their partial defluorination and is in accordance with the observed incomplete fluoride recoveries from the anode.

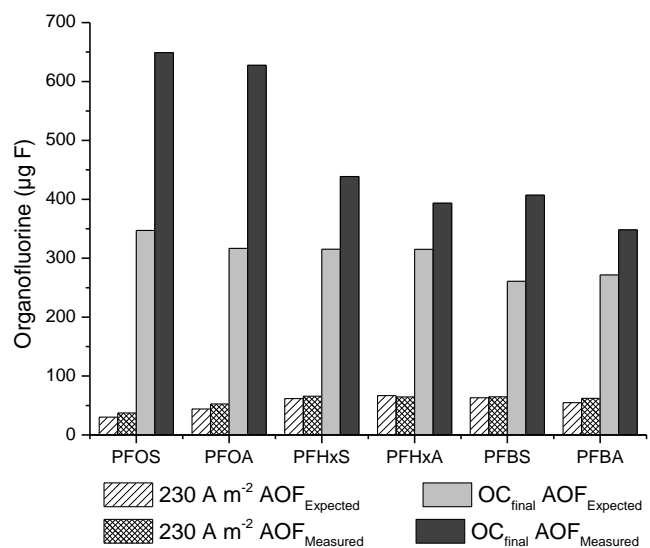

**Figure 3**. AOF values measured in the effluent and AOF values expected from the detected effluent concentrations of PFASs during current application (230 A m$^{-2}$) and in the OC$_{final}$, expressed in µg L$^{-1}$ of F$^-$.

The specific energy consumption for the experiments conducted at 230 A m$^{-2}$ was calculated as 10.1±0.7 kWh m$^{-3}$ (**Text S8**), for reaching PFASs removals in the range 16.6±1.2% (PFBA) - 67±4.1% (PFOS) (**Table 1**). Previous studies employing BDD anodes (Schaefer et al., 2017) and Ti$_4$O$_7$ Magnéli phase anodes (Hui Lin et al., 2018) in similar low conductivity electrolyte solutions reported energy consumption values between 14-46 kWh m$^{-3}$ for the order removal of PFOA and 36-125 kWh m$^{-3}$ for order removal of PFOS. In 100 mM K$_2$HPO$_4$ as supporting electrolyte and using a flow-through Ti$_4$O$_7$ anode, the energy consumption per log removal for PFOA and PFOS was 5.1 and 6.7 kWh m$^{-3}$, respectively (Le et al., 2019). Given the much lower cost of the developed graphene sponge electrode, the reactor performance may be enhanced by using a higher anode surface area, e.g., by stacking up multiple anodes in flow-through mode or using a larger anode than the one employed in this study, thus achieving a more complete electrochemical degradation of PFAS.

**CONCLUSION**

This study demonstrates that graphene-based sponge anode is capable of electrochemically degrading C4-C8 PFCAs and PFSAs in low-conductivity aqueous solution (~1 mS cm$^{-1}$). Electrosorption (7.4-35%) and electrooxidation (9.3-32 %) were determined to be the main removal mechanisms of all PFASs, whereas there was no loss observed due to their volatilization. Increase in the anodic current led to an enhanced removal of the target PFASs, whereas significantly worsened performance at higher flowrates demonstrated that the electrochemical removal of PFASs was mass transfer limited. Fluoride released from the PFASs molecules in the form of HF reacted with the graphene coating and was incorporated into the BRGO anode, likely due to the formation of the covalent C-F, C-F$_2$ and C-F$_3$ bonds. At the highest applied current density (i.e., 230 A m$^{-2}$), electrochemical degradation of both long-chain and short-chain PFASs was confirmed by the near complete recovery of fluoride (i.e., 74%-87 %, relative to the electrooxidized compound) and sulfate (91-98%) in the case of PFOS, PFHxS and PFBS. Thus, although the defluorination efficiencies relative to the overall compound removal were relatively low (i.e., 8-23%), nearly all the electrooxidized fraction of PFASs underwent defluorination and desulfonation. The estimated energy consumption at a current density of 230 A m$^{-2}$ in low conductivity supporting electrolyte (1 mS cm$^{-1}$) was 10.1 ± 0.7 kWh m$^{-3}$.

This proof-of-concept study demonstrated that the developed graphene sponge anode is capable of C-F bond cleavage and due to its demonstrated low electrocatalytic activity towards chloride, without forming toxic chlorinated byproducts. Electrochemical defluorination of PFASs without forming chlorine, chlorate and perchlorate represents a major step towards enabling the treatment of landfill leachate, brine, and other complex, PFAS-laden streams.

## Acknowledgments

The authors would like to acknowledge ERC Starting Grant project ELECTRON4WATER (Three-dimensional nanoelectrochemical systems based on low-cost reduced graphene oxide: the next generation of water treatment systems), project number 714177. ICRA researchers thank funding from CERCA program.

# Supporting Information

# Electrochemical Degradation of Poly- and Perfluoroalkyl Substances (PFAS) using Low-cost Graphene Sponge Electrodes

Nick Duinslaeger[†,‡], Jelena Radjenovic [†,§,*]

†*Catalan Institute for Water Research (ICRA), Emili Grahit 101, 17003 Girona, Spain*

‡*University of Girona, Girona, Spain*

§*Catalan Institution for Research and Advanced Studies (ICREA), Passeig Lluís Companys 23, 08010 Barcelona, Spain*

*\* Corresponding author:*

Jelena Radjenovic, Catalan Institute for Water Research (ICRA), Emili Grahit 101, 17003 Girona, Spain

Phone: + 34 972 18 33 80; Fax: +34 972 18 32 48; E-mail: jradjenovic@icra.cat

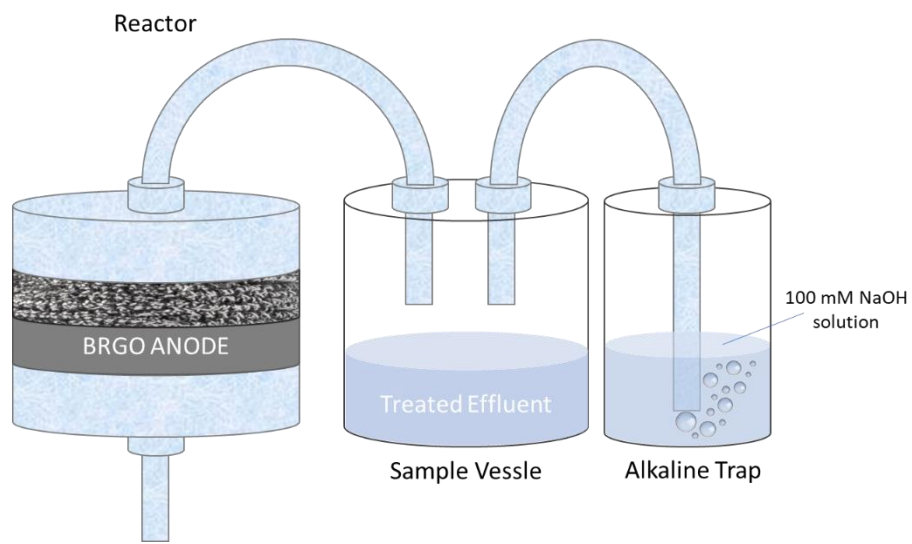

**Figure S1.** Schematic representation of the experimental set-up used to evaluate the loss of the target per- and polyfluoroalkyl substances (PFASs) due to their volatilization.

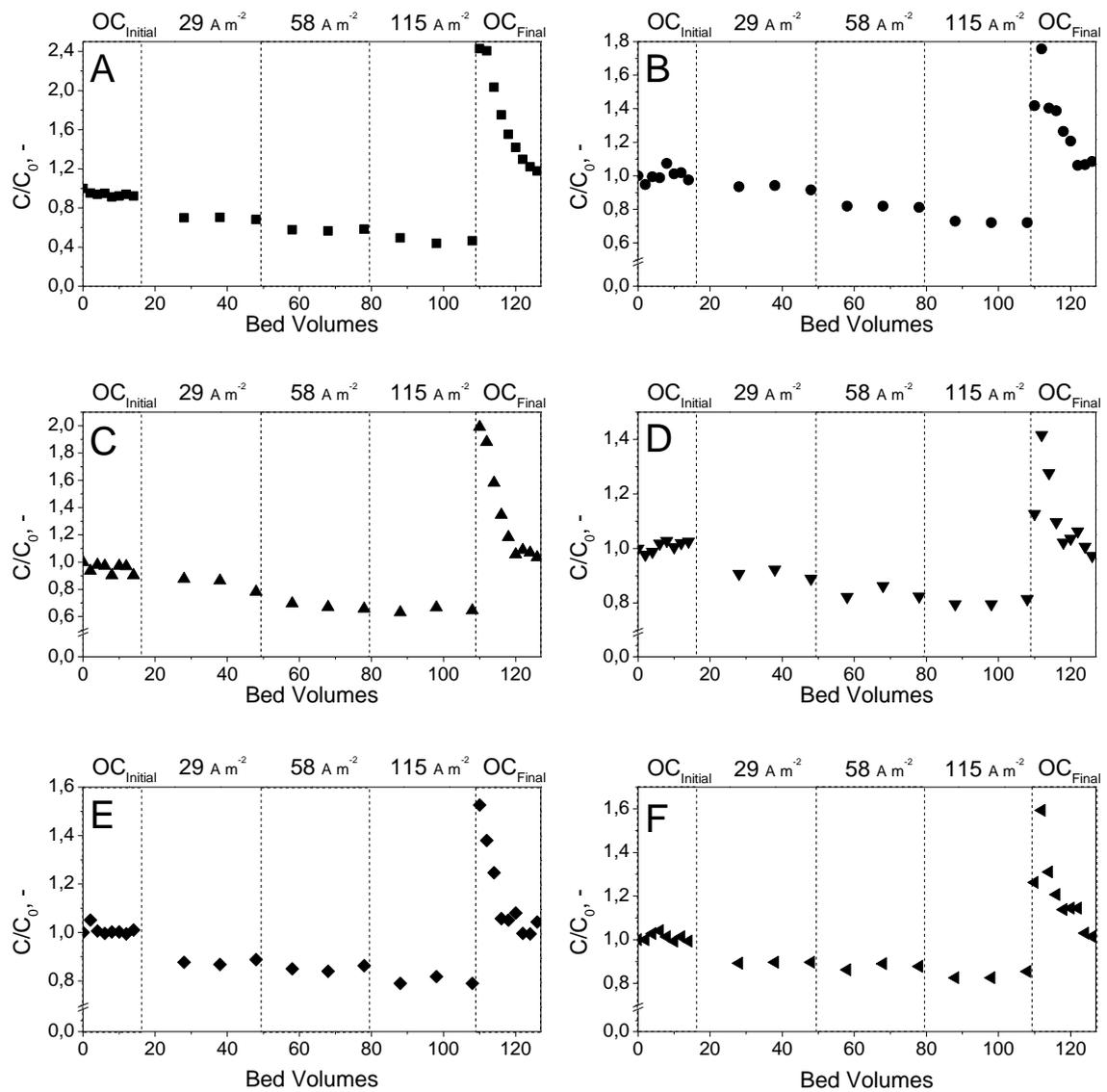

**Figure S2**. Measured concentrations of PFASs normalized to the initial value (C/C$_0$, C$_0$=0.2 µM) at varying anodic currents of 29, 58 and 115 A m$^{-2}$ and effluent flux of 173 LMH: **A)** PFOS, **B)** PFOA, **C)** PFHxS, **D)** PFHxA, **E)** PFBS, and **F)** PFBA. OC$_0$-initial open circuit, OC$_{final}$- final open circuit.

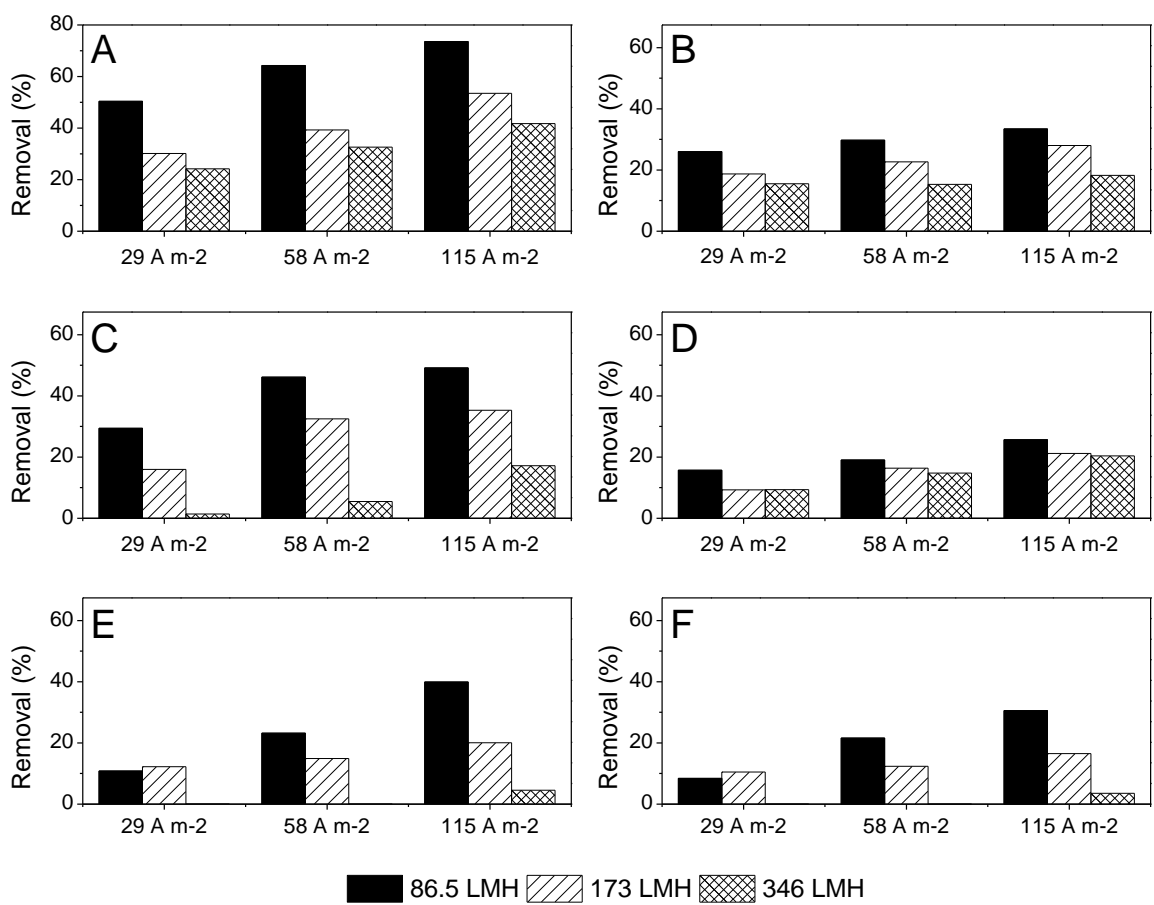

**Figure S3**. Removal efficiencies of: **A)** PFOS, **B)** PFOA, **C)** PFHxS, **D)** PFHxA, **E)** PFBS, and **F)** PFBA at three different effluent fluxes (86.5, 173 and 346 LMH) and anodic current densities of 29, 58 and 115 A m$^{-2}$.

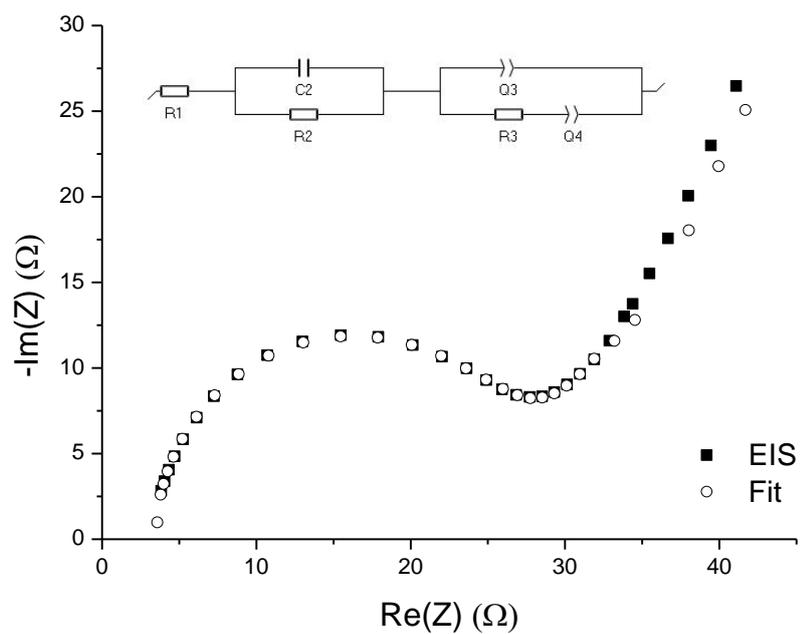

**Figure S4**. Nyquist plot of the impedance, experimental (empty dots) and theoretical spectra (filled dots); the equivalent circuit is presented in the insert; frequency range: 10 000 to 1 Hz.

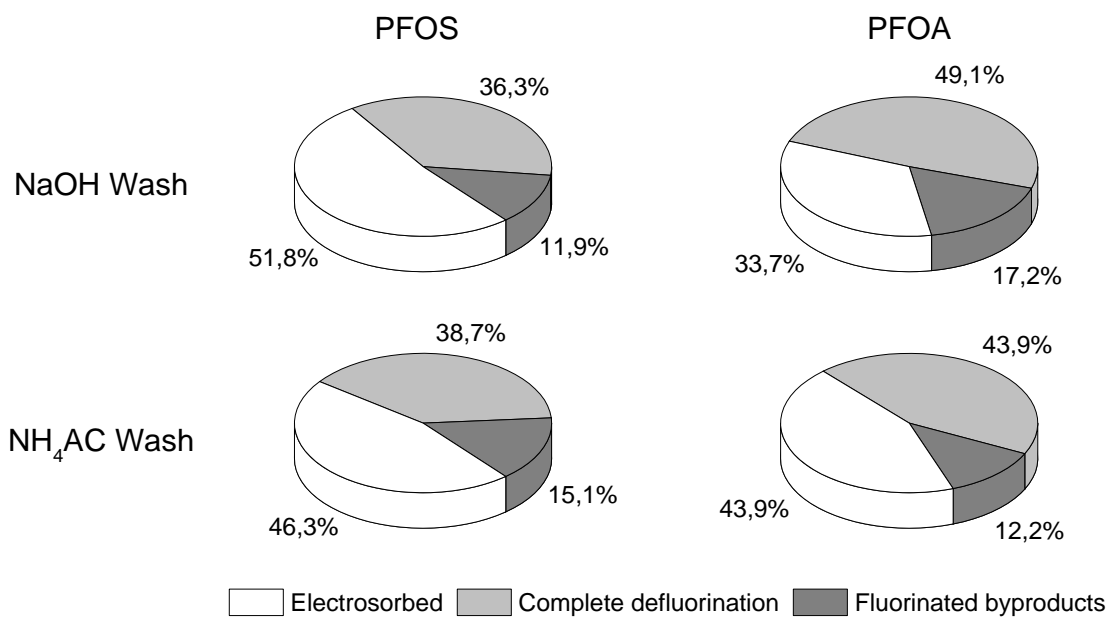

**Figure S5.** Distribution of different removal mechanisms in the overall removal of PFOS and PFOA with washing of the BRGO anode with 100 mM NaOH aqueous solution (top) and 100 mM ammonium acetate in methanol (bottom) before CIC analysis. The experiments were conducted at 230 A m-2 and 2 µM initial concentration of both compounds.

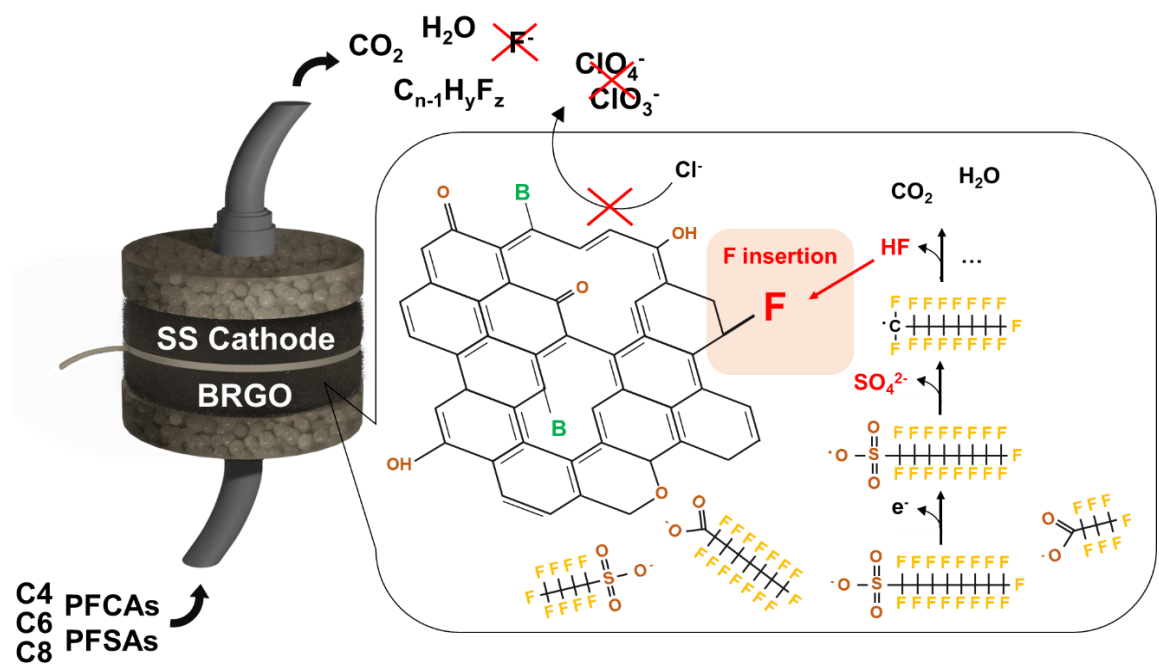

**Figure S6**. Detailed figure of the electro-oxidation of PFAS at the BRGO anode giving way to the incorporation of F on the graphene based structure.

S7

**Table S1.** Summary of the physico-chemical properties of the selected PFASs, including molecular weight (MW, g mol$^{-1}$), theoretical octanol-water partition coefficient (log K$_{OW}$), predicted acid dissociation constant (pKa), and calculated molecular polarizability of the PFAS anion ($a_p$ anion, C m$^2$ V$^{-1}$), as predicted using the DFT modeling [1].

| PFAS | CAS No. | MW (g mol$^{-1}$) | Molecular formula | Structure | Log K$_{ow}$ theoretical value | pK$_a$ theoretical value | $\alpha_p$ anion (C m$^2$ V$^{-1}$) |
|---|---|---|---|---|---|---|---|
| PFBA | 375-22-4 | 214.04 | C$_4$HF$_7$O$_2$ | 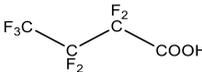 | 2.08-3.93[2–4] | 0.05-1.07[2,5–7] | 77.5 |
| PFHxA | 307-24-4 | 314.05 | C$_6$HF$_{11}$O$_2$ | 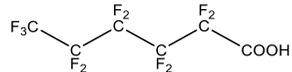 | 2.02-5.97[3,4] | -0.17- 0.84[5,7,8] | 109 |
| PFOA | 335-67-1 | 414.07 | C$_8$HF$_{15}$O$_2$ | 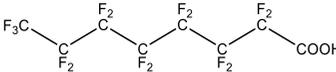 | 2.95-7.75[3,4] | -0.21 – 2.9[5–7] | 140 |
| PFBS | 375-73-5 | | C$_4$HF$_9$O$_3$S | 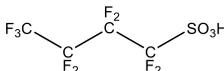 | 0.732-3.68[3,4] | -3.94 – 0.14[5,7,8] | 107 |

| PFHxS | 355-46-4 | 300.10 | $C_6HF_{13}O_3S$ | 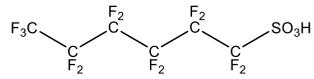 | 2.18-5.25[3,4] | -3.45 – 0.14[5,7,8] | 138 |
| PFOS | 1763-23-1 | 400.11 500.13 | $C_8HF_{17}O_3S$ | 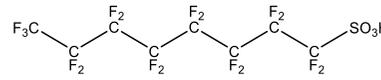 | -0.57 - 7.03[3,4] | -3.41- 0.14[5,7,8] | 170 |

**Table S2** Recoveries of the target PFASs obtained in their adsorbable organic fluorine (AOF) analyses.

| Compound | Recovery (%) |
|----------|--------------|
| PFOS | 59±2.8 |
| PFOA | 75.4±4.2 |
| PFHxS | 93.5±1 |
| PFHxA | 114±4.4 |
| PFBS | 99.5±2.1 |
| PFBA | 147±1.8 |

**Table S3.** The optimized compound-dependent MS parameters: declustering potential (DP), collision energy (CE) and cell exit potential (CXP) for each compound and each transition of the negative mode.

| PFAS  | Q1 Mass (Da) | Q3 Mass (Da) | DP  | CE   | CXP |
|-------|--------------|--------------|-----|------|-----|
| PFBA  | 213.114      | 169.00       | -25 | -14  | -9  |
|       | 213.114      | 119.00       | -25 | -32  | -9  |
| PFBS  | 299.070      | 80.00        | -15 | -66  | -7  |
|       | 299.070      | 98.80        | -15 | -36  | -11 |
| PFHxA | 313.000      | 269.00       | -5  | -12  | -17 |
|       | 313.000      | 119.00       | -25 | -40  | -10 |
| PFHxS | 398.924      | 79.90        | -30 | -74  | -9  |
|       | 398.924      | 99.10        | -30 | -76  | -7  |
| PFOA  | 412.953      | 368.90       | -15 | -14  | -17 |
|       | 412.953      | 169.10       | -15 | -28  | -7  |
| PFOS  | 498.895      | 79.90        | -15 | -104 | -11 |
|       | 498.895      | 98.90        | -15 | -88  | -9  |

**Text S1.** *Characterization of the synthesized graphene-based sponges.*

The characterization of the BRGO sponge is reported in detail in our previous work [9,10]. Below is the summary of the main results reported in these studies.

Scanning electron microscopy (SEM) was performed to evaluate the condition of the RGO coating on the mineral wool template and showed complete coverage of the mineral wool template. The XPS analyses revealed a C/O atomic ratios of 1.7 and 3.5 for GO and BRGO, demonstrating an efficient hydrothermal reduction of GO. The atomic content of the BRGO sponge as determined by the XPS analyses was 75.6% C, 21.8% O, 1.3% B and 1.2% N (N originating from the impurities of the initial GO solution employed in the synthesis). A measure for the content of graphene defects is given by the $I_d/I_g$ ratio, calculated based on the Raman characterization. The $I_d/I_g$ ratio for BRGO was 1.18. The high degree of hydrophobicity of the sponge was confirmed by determining the contact angle of the BRGO sponge. The contact angle determined for BRGO was 139.67°±4.50°, similar to the previously reported values [11]. Next, X-ray diffraction (XRD) analysis showed a decrease in the interlayer spacing from 8.1 Å for GO to ≈ 3.5 Å for BRGO, due to the removal of the oxygen functional groups from the basal plane. The Brunauer–Emmett–Teller (BET) specific surface area was determined 1.39 $m^2$ $g^{-1}$, and the zeta potential of BRGO was –36.2 mV, determined using an aqueous solution at pH 7 using Zetasizer Nano ZS (Malvern Panalytical Ltd) operating with a 633 nm laser.

**Text S2**. *Calculation of the ohmic drop.*

The ohmic drop was calculated from the Ohmic internal resistance obtained in the electrochemical impedance spectroscopy (EIS) experiments of the BRGO – stainless steel system. The EIS experimental data was fitted using the BioLogic EC-lab software and the Nyquist plot and the equivalent circuit of the system for 10 mM phosphate buffer (1 mS cm$^{-1}$) is shown in **Figure S4**.

In the circuit, the ohmic internal resistance (R1) represents the sum of several contributions, which are: electrolyte, separator, current collector and electric conductivity of the active material. R1 is connected in series with two elements that are linked in parallel to each other: the capacitance (C2) and the charge transfer resistance (R2) of the graphene sponge. A constant phase element (Q3), representing the double layer capacitance, which occurs at the interface between the material and the electrolytes due to charge separation and mass transport limitation within the electrodes, is connected in parallel with a charge separation resistance R3, and Q4 that represents the ideal polarizable capacitance. The obtained R1 was 3.4 Ω for 10 mM phosphate buffer resulting in the ohmic drop of 1.4 V for 400 mA of the applied anodic current in 10 mM phosphate buffer. The ohmic drop-corrected anode potential ($E_{corrected}$, V) was calculated according to the following equation:

$$E_{corrected} = E_{recorded} - (I \times R1) \qquad \text{(Eq. S1)}$$

where I is the applied anodic current (A), R1 is the uncompensated resistance (Ω) and $E_{recorded}$ is the anode potential recorded in the chronopotentiometric experiments (V). $\chi^2$ of the fit was 0.194 and represents the criterion for minimization of the fit.

**Text S3.** *UPLC-QqLIT-MS analysis of the target PFASs*

PFOS, PFOA, PFHxS, PFHxA, PFBS, and PFBA were analyzed by a 5500 QTRAP hybrid triple quadrupole-linear ion trap mass spectrometer (QqLIT-MS) with a turbo Ion Spray source (Applied Biosystems, Foster City, CA, USA), coupled to a Waters Acquity Ultra-Performance™ liquid chromatograph (UPLC) (Milford, MA, USA), according to the previously published method [12]. The optimized compound-dependent parameters of the MS are summarized in **Table S2**. Chromatographic separation was accomplished using an Acquity UPLC C18 column (50 mm x 2.1 mm, 1.7 µm). A Phenomenex Luna 5 µm precolumn C8 (50 mm x 3 mm, 100 Å) was employed to minimize the contamination from the mobile phases. The mobile phase used for the chromatographic separation consisted of aqueous ammonium acetate 5 mM (A) and methanol (B) and was delivered at flow rate of 0.4 ml min$^{-1}$. The elution gradient condition started at 10% B and rose to 50% B in 2 min, and then it was linearly increased to 70% B in 4 min, and finally increased to 90% B in 8 min. This percentage was maintained for 1 min more. Finally, the mobile phase was returned to initial conditions in 1 min. Initial conditions were maintained for 1 min more. The sample injection volume was 10 µL. Limit of detection (LOD) for this method was determined to be 0.1 µg L$^{-1}$.

**Text S4.** *Combustion ion chromatography (CIC) analysis*

Determination of adsorbable organic fluorine (AOF) was performed with a combustion ion chromatography (CIC) system for ultra-trace fluorine analysis, consisting of an automated boat controller (ABC-100), an automatic quick furnace (AQF-100) with a water supply unit (WS-100) and a gas absorption unit (GA-100) (all from Mitsubishi Chemical Analytech Co., LTD, Kanagawa, Japan). The combustion unit was linked to an IC system (ICS-2100, Dionex). 100 mL od aqueous sample was extracted on two activated carbon cartridges (Envirosciences, AOX Premium) with the adsorption unit (TXA-04). Then, the adsorbent was transferred quantitatively in a ceramic sample boat and combusted in a furnace at 1,100 ℃, to convert the organically bound fluorine into HF. The HF formed was absorbed in a specific volume of MilliQ water in the adsorption tube ($V_{ads\ tube}$) and measured as fluoride (F⁻) by IC analysis, which enabled its trace level determination (limit of detection (LOD): 0.5 µg L⁻¹). The results were corrected for the blank value obtained by combusting the wetted activated carbon cartridges to obtain the AOF value of the aqueous sample. For quantification of F− by IC, calibration curve was prepared from a sodium fluoride stock solution (1 g L⁻¹). A two-line calibration was established, 0.5, 1, 5, 10, and 25 µg L⁻¹ F⁻ for low concentration range and 25, 50, 100, 200, and 500 µg L⁻¹ F⁻ for high concentration range. To validate the AOF method, the AOF recoveries of the individual PFASs were determined for the initial concentration of 0.2 µM, in 10 mM phosphate buffer, in triplicate and are summarized in **Table S3**.

The maximum expected fluoride concentrations ($C_{F_{exp}}$, µg L⁻¹) from each PFAS, i.e., assuming complete defluorination of each compound, were calculated using Equation S2.

$$C_{F_{exp}} = n_{F_i} \times \frac{MW_F}{MW_{PFAS_i}} \times C_{PFAS_i} \qquad \text{(Eq. S2)}$$

where $n_{F_i}$ is the number of fluorine atoms in the target PFAS, $MW_F$ is the molecular weight of fluorine (19 g mol$^{-1}$), $MW_{PFAS_i}$ is the molecular weight of the target PFAS (g mol$^{-1}$), and $C_{PFAS_i}$ is the initial PFAS concentration (µg L$^{-1}$).

The total amount of AOF ($m_F$, µg of F$^-$) measured in the reactor effluent cumulative samples taken during current application and in the OC$_{final}$ was calculated as follows:

$$m_{AOF} = \sum \left( \frac{V_i \times C_{AOF_i}}{R_{AOF,PFAS}} \right) \qquad \text{(Eq. S3)}$$

where $V_i$ (mL) the volume of sample $i$, $C_{AOF_i}$ (µmol L$^{-1}$) is the fluoride concentration measured in the AOF analysis of the sample $i$, and $R_{AOF,PFAS}$ the recovery of the target PFAS compound as calculated above.

The AOF recovery for each target PFAS ($R_{AOF,PFAS}$ %) was calculated according to the following equation:

$$R_{AOF,PFAS}\,(\%) = \frac{(C_{AOF,masured} - C_{F_{blank}})}{C_{F_{exp}}} \times 100 \qquad \text{(Eq. S4)}$$

where $C_{AOF}$ (µg L$^{-1}$ of F$^-$) is the fluoride concentration measured in the AOF analysis of the target PFAS, $C_{F_{blank}}$ (µg L$^{-1}$) is the fluoride concentration measured in the AOF analysis of the blank sample (i.e., clean extracted water sample), representing the fluoride originating from the activated carbon cartridges; the $C_{F_{blank}}$ was measured in triplicate and was equal to 8±2 µg L$^{-1}$. The results presented are mean of duplicate experiments with their standard deviations (SDs).

Recoveries of each target PFAS in the AOF analysis were determined by analyzing the AOF (i.e., including adsorption and combustion) of standard solutions of single PFASs (0.2 µM) in 10 mM phosphate buffer with AOF analysis according to the procedure explained in **Text S3**. The general trend was that higher recoveries in the AOF analysis were observed for PFCAs compared with PFSAs, and that higher recoveries are observed with the shortening of the fluorinated chain length, i.e., the recoveries were increasing in the order PFOS<PFHxS<PFBS, and PFOA<PFHxA<PFBA (**Table S2**). Cleavage of the C7-C8 bond in PFOA was found to be thermodynamically more favorable than defluorination at temperatures below 1000 °C [13,14]. Although the combustion in the AOF analysis was conducted at 1,100 ℃ (**Text S4**), it may not be complete and the formation of shorter chain fluorinated byproducts cannot be excluded, thus leading to lower AOF recoveries for longer chain PFOA and PFOS. The impact of the terminal functional group (i.e., carboxylic and sulfonic) on the AOF recovery can be attributed to the strength of the interactions between PFAS molecules and the surface of the activated carbon cartridge employed in the AOF analysis. According to Baghirzade et al. [15], the initial desorption of the PFAS loaded onto the activated carbon (prior to its transition into the gas phase) is governed by the molecule´s polarizability. Van der Waals interactions induced by PFAS with higher polarizability are stronger[16] and thus may facilitate greater retention of the longer-chain PFASs at the activated carbon cartridge, resulting in their lower volatilization in the combustion chamber of the CIC instrument and thus lower recoveries of the AOF analysis[17,18]. Once desorbed, thermolysis of PFASs will take several steps to achieve the PFASs decomposition and defluorination during combustion. Also, lower recoveries were consistently obtained for PFSAs compared with PFCAs of the same chain length. Also, for two compounds, PFHxA and PFBA, recoveries higher than 100% were obtained (i.e., 114±4.4% and 147±1.8%, respectively). Overall recoveries higher than 100% have been previously reported in the AOF analysis of short-chain PFASs [19,20], yet it is unclear what is causing the interference. It should be noted that all analysis were corrected for the baseline fluoride concentration measured for the activated carbon cartridges used in the analysis. IC analysis did not show any interference in the fluoride peak measured in the AOF analysis of PFHxA, PFBA or other PFASs, yet it is possible that the combustion of these compounds adsorbed onto activated carbon cartridges may have led to the formation of byproducts that co-elute with $F^-$ ion. In addition, the recoveries of AOF analysis will be highly dependent on the combustion tube material and details of the analysis (e.g., combustion temperature, carrier gas flowrate)[21].

**Text S5.** *Determination of PFASs electrosorption and electrooxidation*

The removal of the target PFASs via electrosorption was calculated as follows:

$$Electrosorption\ (\%) = \frac{\sum(C_{OCfinal_i} \times V_{OCfinal_i})}{\sum(M_{CA_j} \times V_{CA_j})} \times 100 \qquad (Eq.\ S5)$$

where $c_{OCfinal,i}$ is the molar concentration (µmol L$^{-1}$) of the compound in the sample $i$, $V_{OCfinal_i}$ the volume of sample $i$ and $M_{CA_j}$ the molar concentration (µmol L$^{-1}$) of the compound in sample $j$, $V_{CA_j}$ the volume of sample $j$, with i representing the samples taken in the final OC and j the samples during current application.

The removal of the target PFAS via electrooxidation was calculated as follows:

$$Electrooxidation\ (\%) = \frac{\sum(M_{CA_j} \times V_{CA_j}) - \sum(M_{OCfinal_i} \times V_{OCfinal_i})}{\sum(M_{CA_j} \times V_{CA_j})} \times 100 \qquad (Eq.\ S6)$$

**Text S6.** *Sulfate recovery from PFASs*

The recovery of sulfate released in the electrooxidation of the target PFASs was calculated as:

$$Sulfate\ recovery\ (\%) = \frac{\Sigma(C_{SO_4^{2-}} \times V)}{N_{PFAS}} \times 100 \qquad (Eq.\ S7)$$

where $C_{SO_4^{2-}}$ the measured sulfate concentration (µg L$^{-1}$) in the recovery solution used for sulfate desorption, $V$ is the volume of the recovery solution, $N_{PFAS}$ the calculated total amount of PFAS$_i$ electrooxidated (µmol) in that experiment.

**Text S7.** *Fluoride recovery from PFASs*

The recovery of fluoride released in the electrooxidation of the target PFASs and analysed using CIC of the graphene sponge anode was calculated as:

$$Fluoride\ recovery\ (\%) = \frac{(C_{F_{sponge}} - C_{F_{MW}}) \times V_{ads\ tube} \times MW_F}{N_{PFAS} \times n_F} \times 100 \qquad (Eq.\ S8)$$

where $C_{F_{sponge}}$ the fluoride concentration (µg L$^{-1}$) measured in the CIC analysis of the graphene sponge anode after each PFAS electrooxidation experiment, $C_{F_{MW}}$ is the fluoride concentration (µg L$^{-1}$) measured in the CIC analysis of the blank mineral wool, $V_{ads\ tube}$ is the total volume of the Milli Q water in the adsorption tube of the CIC, $N_{PFAS}$ is the amount of PFAS degraded during electrooxidation (µmol) and $n_F$ is the number of fluorine atoms in the target PFAS, MW$_F$ is the molecular weight of fluoride (19 g mol$^{-1}$).

**Text S8.** *Energy consumption*

The energy consumption ($E$, kWh m$^{-3}$) was calculated as follows:

$$E = \frac{V_{cell} \times I}{Q} \quad \text{(Eq. S9)}$$

where $I$ is the applied anodic current (A), $V_{cell}$ is the cell potential (V), and $Q$ is the reactor flow rate (m$^3$ h$^{-1}$).